\documentclass[%
reprint,
superscriptaddress,
amsmath,amssymb,
aps,
prl,
]{revtex4-2}

\usepackage{graphicx}% Include figure files
\usepackage{dcolumn}% Align table columns on decimal point
\usepackage{xcolor}
\usepackage{ulem} % Required for strike-through command \sout{}
\usepackage{bm}% bold math
\usepackage{hyperref}% add hypertext capabilities
\usepackage{float}
\newcommand{\iitj}{\affiliation{Department of Physics, Indian Institute of Technology Jodhpur\\
		N.H. 62, Nagaur Road, Karwar, Jodhpur, Rajasthan, India - 342030. }}
\newcommand{\sut}{\affiliation{Department of Mathematics, School of Science, Computing and Engineering Technologies, Swinburne University of Technology, Melbourne, Victoria 3122, Australia}}% Force line breaks with \\
\begin{document}
	
	%\begin{widetext}
	%\onecolumngrid
	%\textcolor{red}{Strikethrough text in red color indicates deleted text in the revised version.}\\
	%\textcolor{blue}{The blue-colored text indicates newly added or modified text in the revised manuscript.}
	%\end{widetext}
	%\twocolumngrid
	\title{Salt Effects on Ionic Conductivity Mechanisms in Ethylene Carbonate Electrolytes: Interplay of Viscosity and Ion-ion Relaxations}
	\newcommand{\revsapta}[1]{{\color{red}{#1}}}
	\preprint{APS/123-QED}
	
	\author{Hema Teherpuria}
	\iitj
	\author{Sapta Sindhu Paul Chowdhury}
	\iitj
	\author{Sridhar Kumar Kannam}
	\sut
	\author{Prabhat K. Jaiswal}
	\iitj
	\author{Santosh Mogurampelly}%
	\email{santosh@iitj.ac.in}
	\iitj
	
	
	%\affiliation{
	%\affiliation{
	%\affiliation{%
	
	
	\date{\today}% It is always \today, today,
	
	\begin{abstract}
		The intricate role of shear viscosity and ion-pair relaxations on ionic conductivity mechanisms and the underlying changes induced by salt concentration ($c$) in organic liquid electrolytes remain poorly understood despite their widespread technological importance. Using molecular dynamics simulations employing nonpolarizable force fields for $c$ ranging between 10$^{-3}$ to 10$^1$ M, we show that the low and high $c$ regimes of the EC-LiTFSI electrolytes are distinctly characterized by $\eta\sim\tau_c^{1/2}$ and $\eta\sim\tau_c^{1}$, where $\eta$ and $\tau_c$ are shear viscosity and cation-anion relaxation timescales, respectively. Our extensive simulations and analyses suggest a universal relationship between the ionic conductivity and $c$ as $\sigma(c)\sim c^{\alpha}e^{-c/c_{0}} (\alpha>0)$. The proposed relationship convincingly explains the ionic conductivity over a wide range of $c$, where the term $c^\alpha$ accounts for the uncorrelated motion of ions and $e^{-c/c_0}$ captures the salt-induced changes in shear viscosity. Our simulations suggest vehicular mechanism to be dominant at low $c$ regime which transitions into a structural diffusion mechanism at high $c$ regime, where structural relaxation is the dominant form of ion transport mechanism. Our findings shed light on some of the fundamental aspects of the ion conductivity mechanisms in liquid electrolytes, offering insights into optimizing the ion transport in EC-LiTFSI electrolytes.
		
		%\begin{description}
		%\end{description}
	\end{abstract}

\maketitle

%\tableofcontents

%\\ %The line
%break was forced \lowercase{via} \textbackslash\textbackslash

The ion transport mechanisms are well understood for different classes of polymer electrolytes but the effects arising from salt concentration in liquid electrolytes and the intricate role of shear viscosity and ion-pair relaxations are not well established.\cite{borodin2006,maitra2007,balsara2018,ganesan2019} Typically, the ionic conductivity of liquid electrolytes is significantly higher than that of solid polymer electrolytes at the same salt concentration. In technological applications such as the traditional lithium ion batteries, liquid electrolytes offer huge advantage with high ionic conductivity where lithium-based salts are dissolved in organic solvents along with a variety of molecular compounds.\cite{santosh2016} Understanding the underlying ionic conductivity mechanisms at a fundamental level is critically important since the performance of an electrolyte in rechargeable battery applications depends on the interaction between available charge carriers and their capability to conduct ionic charge at a certain salt concentration.\cite{cai2020} Therefore, it is essential to deeply understand the ion-ion relaxation phenomenon, the effect of salt concentration on shear viscosity, and the connection between them with implications to the ionic conductivity.

The ion transport mechanism in polymer electrolytes has been extensively explored through both experimental and theoretical approaches\cite{maitra2007,borodin2000,borodinpoly2006,diddens2010,santosh2016}. In polyethylene oxide electrolytes, lithium ions are situated at specific coordinate sites near ether oxygen (EO) groups in polymer chains, undergoing continuous segmental motion. Consequently, lithium ions traverse from one EO site to the other along the backbone of polymer chain and intermittently jump between chains because of the segmental motion of polymeric chains. The diffusion of ions and the ionic conductivities are intricately linked to polymer dynamics, manifesting as ion motion through structural relaxation of the polymer matrix or an ion hopping mechanism, where ions hop along the polymer backbone\cite{maitra2007,borodin2000,borodinpoly2006,jordan2017}.

In a recent experimental work, Mongcopa et al.\cite{mongcopa2018} proposed an interesting mechanism for the origin of ionic conductivity in poly(ethylene oxide) lithium bis-(trifluoromethane) sulfonimide (PEO-LiTFSI) electrolytes over a wide range of salt concentrations, by arguing the importance of polymer friction coefficient as a manifestation of the polymer segmental motion at monomeric level. However, for liquid electrolytes, an analogous description in terms of the friction coefficient is not possible, and a direct analysis of the viscosity and its connection to the ionic conductivity is a suitable approach. A few experimental groups investigated the salt concentration effects on organic solvent based liquid electrolytes, but molecular level attempts using computational studies examining the intricate role of shear viscosity and ion-pair relaxations on ionic conductivity are limited\cite{nilsson2020,moumitapccp2023}. Inspired by the experimental study of Mongcopa et al.\cite{mongcopa2018}, we ask how the ionic conductivity is dictated by viscosity over a wide range of $c$ spanning three orders of magnitude. We examine if there is a universally applicable relationship between $\sigma$ and $c$ that predicts the ionic conductivity across the entire spectrum of salt concentrations.

We used classical molecular dynamics (MD) simulations containing intra- and intermolecular interaction terms, \textit{viz}., harmonic potentials for bonds and angles, Fourier terms for torsions, Lennard-Jones and Coulomb potentials for nonbonded interactions. A real space cutoff of 12 Å is employed for nonbonded interactions and k-space summation for the electrostatic interactions is carried out using the particle mesh Ewald method\cite{tom1993}. The force field parameters are obtained from the nonpolarizable OPLS set\cite{jorgensen1996} with geometric combination rules enforced for the cross terms. Nonpolarizable interaction potential models with full charges on ionic species yield inconsistent results with experiments, with decreased ion transport properties and increased densities and viscosities. To address this, we have scaled the partial charges on ionic species to 0.8$e$, offering a judicious choice over computationally more challenging polarizable models or \textit{ab-initio} MD simulations. Using the charge scaling method as a mean-field like approach to treating the induced polarization effects within classical level atomistic simulations, as routinely employed in the literature of electrolytes,\cite{moumitapccp2023,wrobel2021} we obtained transport and structural properties reasonably comparable to experiments\cite{martinez1999}.(see Table S3) We prepared different EC-LiTFSI electrolyte systems with $c$ ranging from $10^{-3}$ to 10$^1$ M at a temperature of 323 K. Long production runs were conducted in NPT ensemble, \cite{md1984,cd1985,bussi2007} with periodic boundary conditions in all three directions and the simulations are performed using the \textsc{GROMACS} 2021 software\cite{abraham2015}. Equilibrium MD simulation trajectories are used for calculating the physical quantities such as diffusion coefficient, viscosity (Green-Kubo method)\cite{hess2002}, ion-pair relaxation time, and the Nernst-Einstein (NE) conductivity\cite{frenkel2002,hansen2006,ciccotti1976}. Nonequilibrium MD (NEMD) simulations with an externally applied electric field are used for calculating the total ionic conductivity to capture ion-ion correlation effects\cite{frenkel2002}.

\begin{figure}
	\includegraphics[height=6.5cm,keepaspectratio]{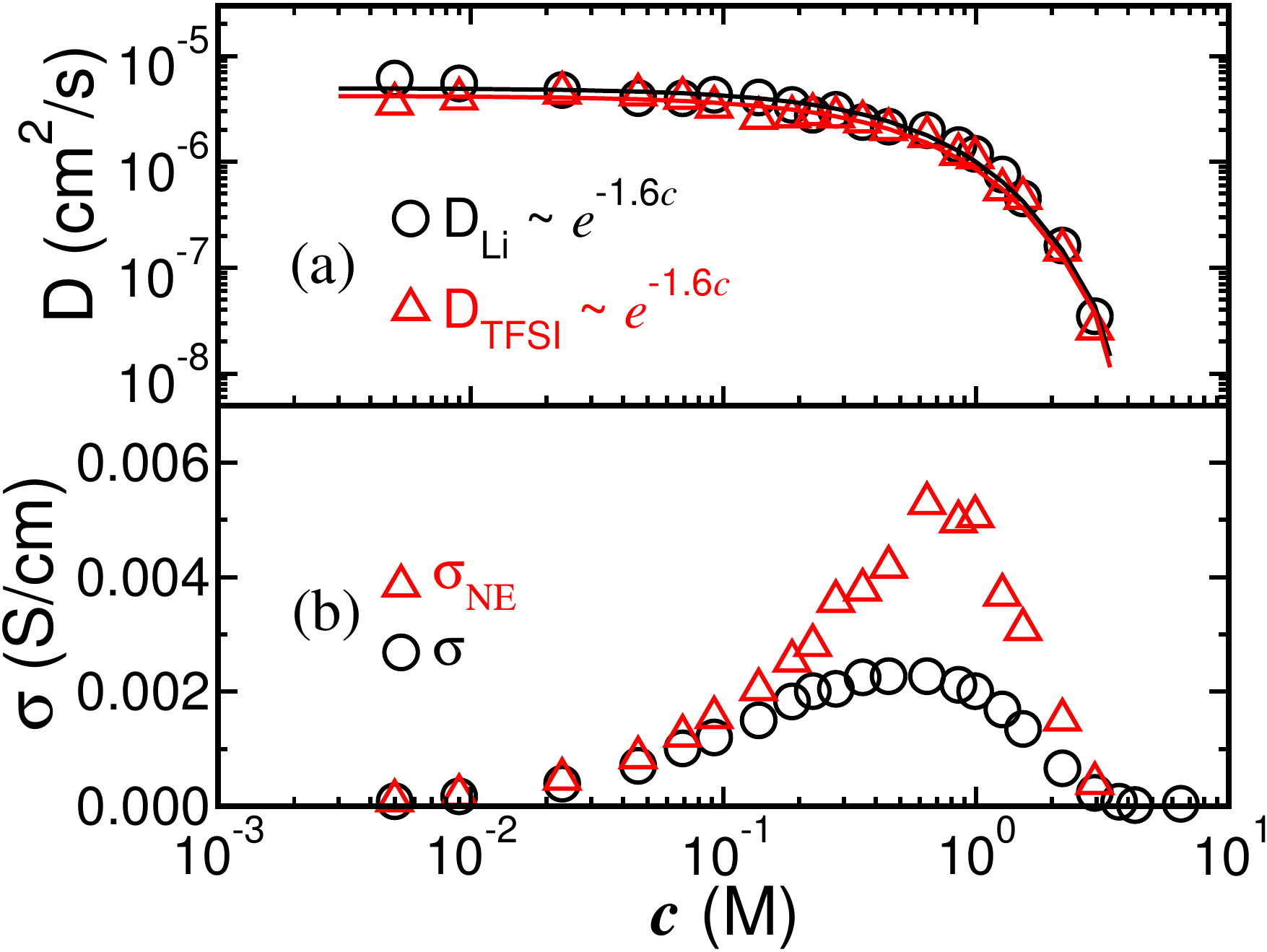}% Here is how to import EPS art
	\caption{(a) Diffusion coefficient of Li$^+$ and TFSI$^-$ ions and (b) The uncorrelated ionic conductivity ($\sigma_{NE}$) calculated using the Nernst-Einstein approximation and the total ionic conductivity ($\sigma$) calculated by applying an external electric field within the NEMD formalism.}
		\label{fig:figure1}
	\end{figure}

Fig. \ref{fig:figure1}(a) demonstrates a mere 2\% difference in diffusivity between Li$^+$ and TFSI$^-$ ions, implying their similar diffusion behaviors at a given salt concentration, corroborating the observations from Wrobel et al.\cite{wrobel2021} and Devaux et al.\cite{devaux2012} Interestingly, we observed a moderate impact of salt concentration on the diffusion coefficient up to 0.5 M, beyond which a rapid and monotonic decrease was evident. At higher salt concentrations, there is a notable reduction in diffusivities, showcasing a 100-fold decrease when the salt concentration increases from 0.25 M to 2.5 M. Consequently, the ions exhibit increased immobility at these elevated salt concentrations.

The overall trend of the diffusion coefficient fits nicely to an exponential decay, yielding equations for $D_{Li}^+$ as 4.97 $\times 10^{-6}  e^{-1.6c}$ and for D$_{TFSI}^-$ as = 4.20 $\times 10^{-6}e^{-1.6c}$. The decrease in diffusivity at higher salt concentrations can be attributed to the enhanced system density. Beyond $\sim$ 0.1 M, the coordination shells of ionic species become more closely packed, limiting the diffusing pathways for ion movement within the electrolyte matrix. This leads to an exponential reduction in ion diffusivity, resembling the findings observed in the diffusion of Li$^+$ and PF$_6^-$ ions within concentrated LiPF$_6$ in propylene carbonate solutions\cite{gustav2021}.

The ionic conductivity results presented in Fig. $\ref{fig:figure1}$(b) clearly reveal that both $\sigma_{NE}$ and $\sigma$ exhibit similar qualitative behavior regarding their dependency on salt concentration. Due to the absence of effects related to the correlated motion of ions\cite{sasaki2023} in the Nernst-Einstein approximation, $\sigma_{NE}=\frac{e^2}{Vk_BT} \left[N_{Li} z_{Li}^2 D_{Li}+N_{TFSI} z_{TFSI}^2 D_{TFSI} \right]$, the value of $\sigma_{NE}$ is consistently higher than $\sigma$ across all $c$ values. Unlike the diffusion coefficient, the behavior of ionic conductivity demonstrates distinctive trends at low, intermediate, and high $c$ values. At lower $c$ values, we observed a monotonic increase in ionic conductivity with salt concentration, contrary to the trends observed for the diffusion coefficient. This increase in ionic conductivity occurs because the number of ionic charge carriers in the EC-LiTFSI electrolyte steadily increases with salt concentration,\cite{ravikumar2018,fuoss1933} while the influence on ionic diffusivities remains marginal in the same $c$ region. Furthermore, it is noteworthy that at low $c$, the ion-ion correlations are minimal due to the high dilution, causing the total ionic conductivity to closely agree with the data predicted by the Nernst-Einstein approximation\cite{ciccotti1976}.

After the initial increase, ionic conductivity slows down with $c$ and reaches a maximum around $c=0.6$ M. The maximum value of total ionic conductivity calculated with the NEMD approach is found to be $2.27\times 10^{-3}$ S/cm, in a reasonable agreement with previous experimental works\cite{nilsson2020,kondo2000}. Beyond the peak, a further increase in $c$ resulted in a significant drop in $\sigma$. The slowing down of $\sigma$ before peaking and the subsequent decrease after reaching maximum are due to the detrimental effects arising from the increasingly dominant prevalence of the cation-anion correlations at high $c$\cite{mongcopa2018}. The rapid decay observed for the ionic conductivity is also connected to the formation of large-sized ionic clusters in the electrolytes at high salt concentrations,(see Fig. S10) which lowers the number of free ionic charge carriers and consequently have detrimental effects on the overall ionic conductivity \cite{mongcopa2018,ding2017}. Overall, the above results suggest that the behavior of Li$^+$ and TFSI$^-$ ions in EC-LiTFSI electrolytes is strongly related to the salt concentration.

\begin{figure}
	\includegraphics[height=6.5cm,keepaspectratio]{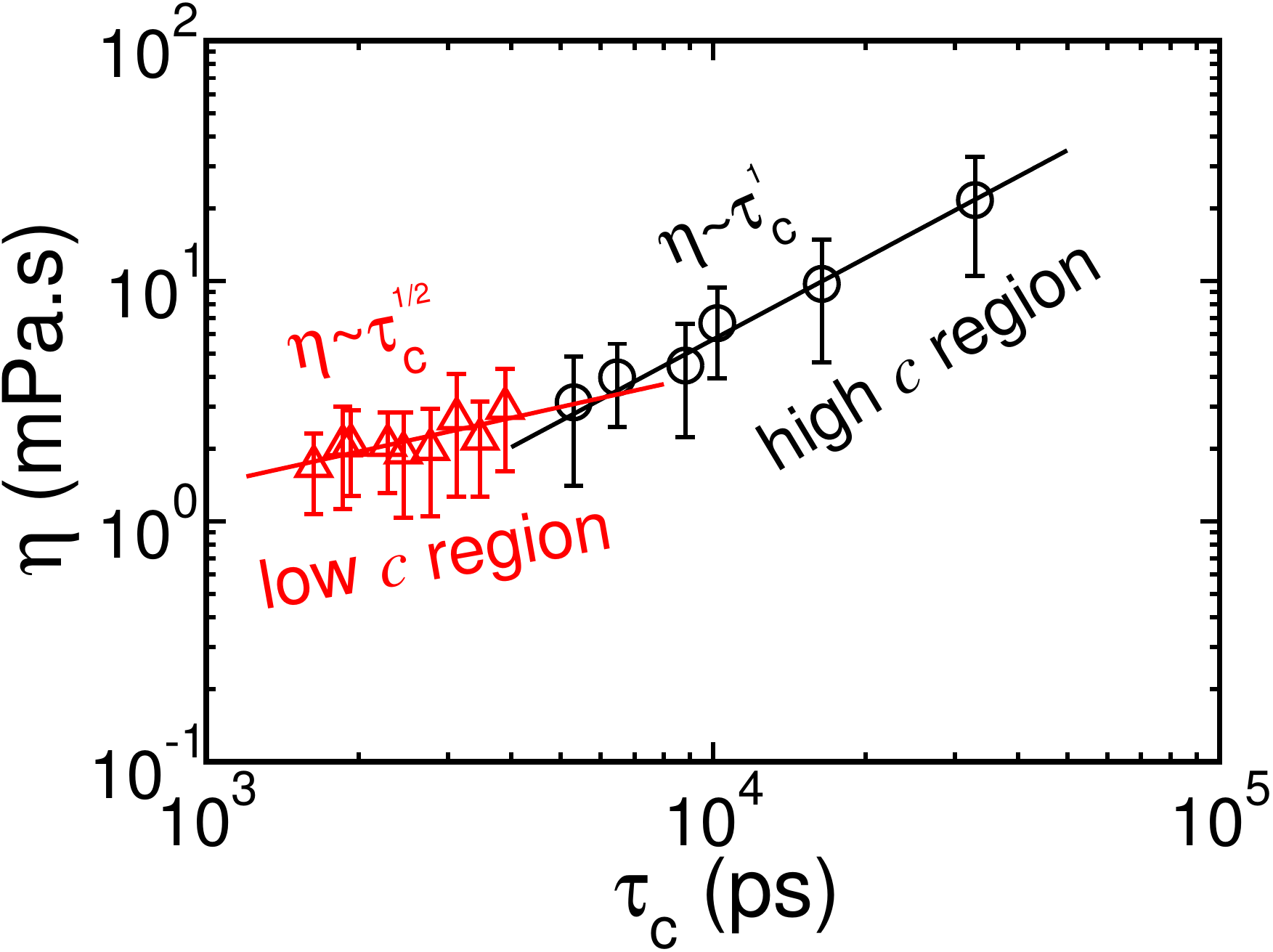}% Here is how to import EPS art
	\caption{Comparison of the shear viscosity against the ion-pair relaxation timescales, yielding $\eta\sim\tau_c^{1/2}$ at low $c$ regime and $\eta\sim\tau_c^{1}$ at high $c$ regime.}
	\label{fig:figure2}
\end{figure}

In the context of glassy systems, there exists a proposition suggesting that an underlying relaxation time dictating transport properties bears a resemblance to viscosity\cite{vishwas2014}. This implies a clear link between the relaxation time and viscosity, where their relationship follows a power-law pattern, resulting in a unity power-law exponent. Our analysis presented in Fig. \ref{fig:figure2} reveals that $\tau_c$ and $\eta$ exhibit two distinct relationships within the low and high $c$ regimes. A power-law fit to the simulation data, when the exponent is kept as a free parameter, yields $\eta = 5.6\times 10^{-2}$ $\tau_c^{0.47}$ in the low $c$ regime and $\eta=1.79\times 10^{-4}$ $\tau_c^{1.13}$ in the high $c$ regime. Subjected to numerical uncertainties in the exponents, the aforementioned findings emphasize the presence of two distinct low and high $c$ regimes. In the low $c$ regime, the behavior is characterized by $\eta\sim\tau_c^{1/2}$, whereas in the high $c$ regime, it is characterized by $\eta\sim\tau_c^{1}$. These findings align internally with the observed discrepancies in the dependency of $\eta$ and $\tau_c$ on $c$. The one-to-one correlation found between $\eta$ and $\tau_c$ in the high $c$ regime is intriguing and carries significant computational implications (i.e., computationally involving $\eta$ calculations can be circumvented and rely simply on $\tau_c$ which is much easier to calculate).

\begin{figure}
	\includegraphics[height=6cm,keepaspectratio]{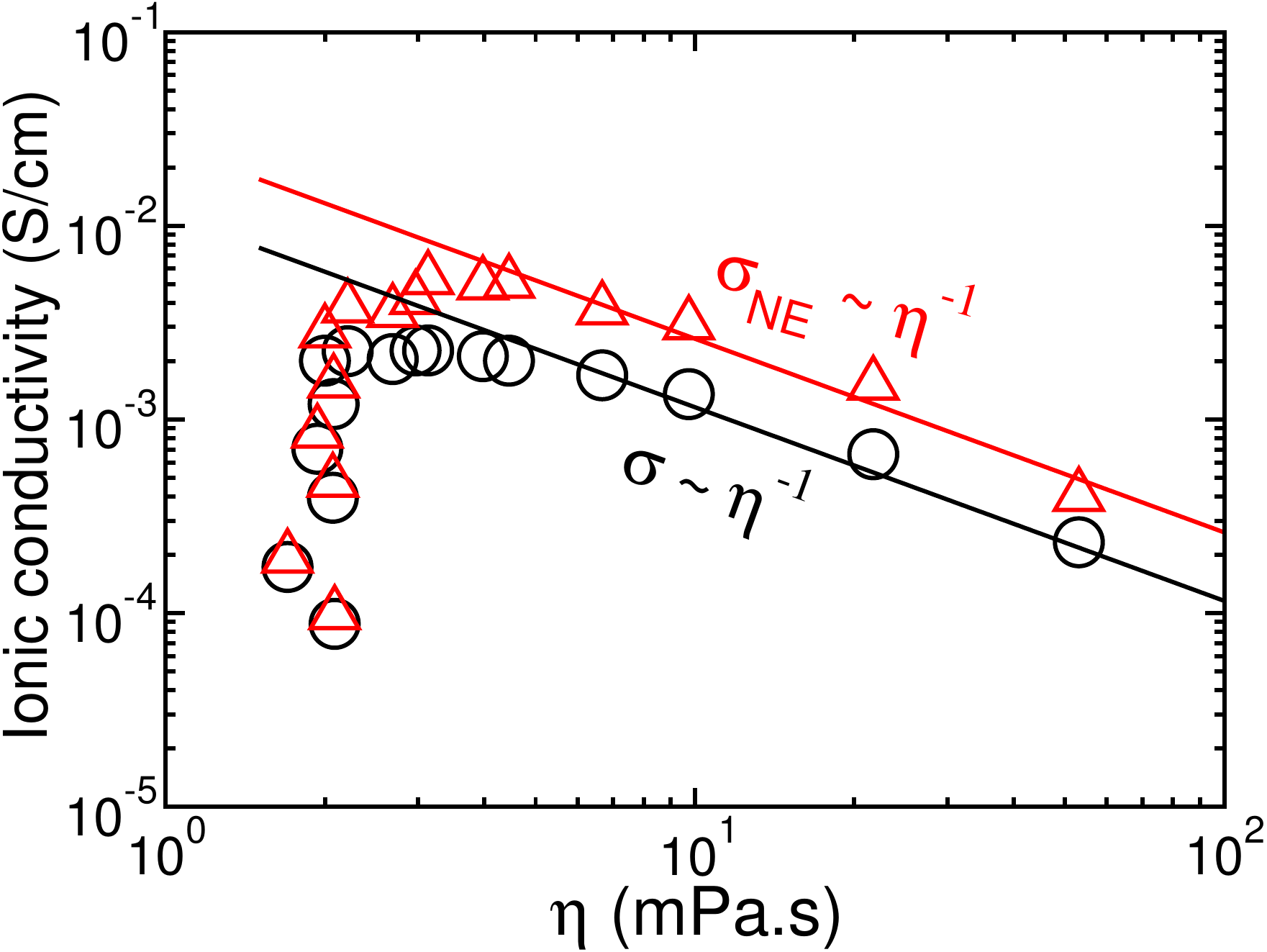}% Here is how to import EPS art
	\caption{Ionic conductivity as a function of viscosity in EC-LiTFSI electrolytes. The low $c$ regime of ionic conductivity is clearly independent of the shear viscosity.}
	\label{fig:figure3}
\end{figure}

Noting that the diffusivities and ionic conductivity vary quite distinctly with the $c$, we anticipate that the power-law relationship governing ionic conductivity, i.e., $\sigma_{NE}=\alpha$/$\eta^{\lambda}$ and $\sigma$ = $\alpha$/$\eta^{\lambda}$, should differ notably from the respective relationships for ionic diffusivities. Accordingly, both the Nernst-Einstein and the total ionic conductivity showed nonmonotonic dependency when plotted against $\eta$ (see Fig. \ref{fig:figure3}). We find that the conductivity is largely insensitive to the changes induced by salt on $\eta$ at low $c$. This result concludes that neither the viscosity nor the ion-pair relaxation phenomena explain conclusively the ionic conductivity at low $c$. Therefore, since no significant insights are gained through the analysis of viscosity and ion-pair relaxations (because $\eta\sim\tau_c^{1/2}$), we conclude that $\sigma$ follows $\sigma \sim c$ at low $c$ due to the increased availability of uncorrelated ionic species. However, we hypothesize that ion-ion correlations still influence $\sigma$ even at low concentrations, modifying the relation to $\sigma \sim c^{\alpha}$ with $\alpha > 0$.

At high $c$, the ionic conductivity decreases rapidly with both the $\tau_c$ and $\eta$, in qualitatively similar manner. Because the shear viscosity increases exponentially at high $c$, we fitted the simulation data to power-law relations at high $c$ as $\sigma =\alpha$/$\eta^{\lambda}$, to understand the degree of correlations between (i) $\sigma_{NE}$ and $\eta$ and (ii) $\sigma$ and $\eta$. We found that the ionic conductivity correlates less sensitively to the ion-pair relaxation times as $\sigma_{NE}\sim\tau_c^{-0.73}$ and $\sigma\sim\tau_c^{-0.76}$. These weak correlations arise from the limitations of $\tau_c$ in fully capturing the relaxation dynamics of different ion associations at high $c$, including isolated ion-pairs, multiplets, and clusters.\cite{siprajcp2023,siprananoscale2024} Similar analysis demonstrated that the Nernst-Einstein and total ionic conductivity are excellently correlated to the viscosity, as $\sigma_{NE}\sim\eta^{-1}$ and $\sigma\sim\eta^{-1}$, at high $c$ regimes. This correlation arises because the ratio of $\sigma$ to $\sigma_{NE}$ remains constant at high $c$, indicating that increased viscosity restricts ion mobility and results in both $\sigma$ and $\sigma_{NE}$ exhibiting similar inverse dependencies on $\eta$ and $\tau_c$.

So far, we have analyzed the mechanisms of diffusion and ionic conductivity by invoking the salt-induced changes in viscosity and ionic conductivity. Because of the monotonic behavior of $D,  \eta$, and $\tau_c$ on $c$, it has been clearly established that the salt concentration effects on diffusivity can be directly interpreted as $D\sim$ $e^{-1.6c}$ for the entire range of $c$. However, due to the nonmonotonic behavior of $\sigma$ and monotonic behavior of $\eta$ and $\tau_c$ on $c$, we have arrived at two distinct salt concentration regimes and find a decent relationship between $\sigma$ and $c$ for low and high $c$ regimes.

Inspired by the work of Mongcopa et al., \cite{mongcopa2018} we propose a unified relationship between the $\sigma$ and c that holds true for the entire range of $c$. Mongcopa et al.\cite{mongcopa2018} proposed that ionic conductivity in PEO-LiTFSI can be explained for the whole salt concentration region by invoking the analysis of the polymer friction coefficient. However, for the liquid electrolytes, the friction coefficient is irrelevant and require the direct analysis of viscosity which we have discussed in the previous section. Our analysis revealed that $\eta$ scales as $e^{c/c_0}$, implying that $\sigma \sim e^{-c/c_0}$ at high $c$, similar to the role of $\xi$ in polymer electrolytes\cite{mongcopa2018}.

Further, to explain the ionic conductivity data across a wide range of salt concentrations including the low and high $c$ regimes, we propose a unified equation as $\sigma=k$$c^{\alpha}e^{-c/c_{0}}$, where $k$ is a constant. The fits of simulation data to the above equation yields $\alpha=0.88$, $c_{0}=0.64$. The fitting of simulation data to the proposed unified eq. $\sigma=k$$c^{\alpha}e^{-c/c_{0}}$ resulted in the location of maximum at 0.54 M, in agreement with the analytical result $\alpha{c}_{0}=0.56$ M. Similarly, the maximum of ionic conductivity is 0.0025 S/cm occurring at a salt concentration of  $\alpha{c}_{0}$. The proposed unified equation fits to the simulation data excellently at all the ranges of $c$. Considering the differences between Mongcopa et al. \cite{mongcopa2018} and this work, such as polymer vs. liquid electrolyte, the unified eq. $\sigma=k$$c^{\alpha}e^{-c/c_{0}}$ holds promise in unraveling the ion conductivity mechanisms in a variety of liquid and polymer electrolytes (see Fig. S14).

\begin{figure}
	\includegraphics[height=6.5cm,keepaspectratio]{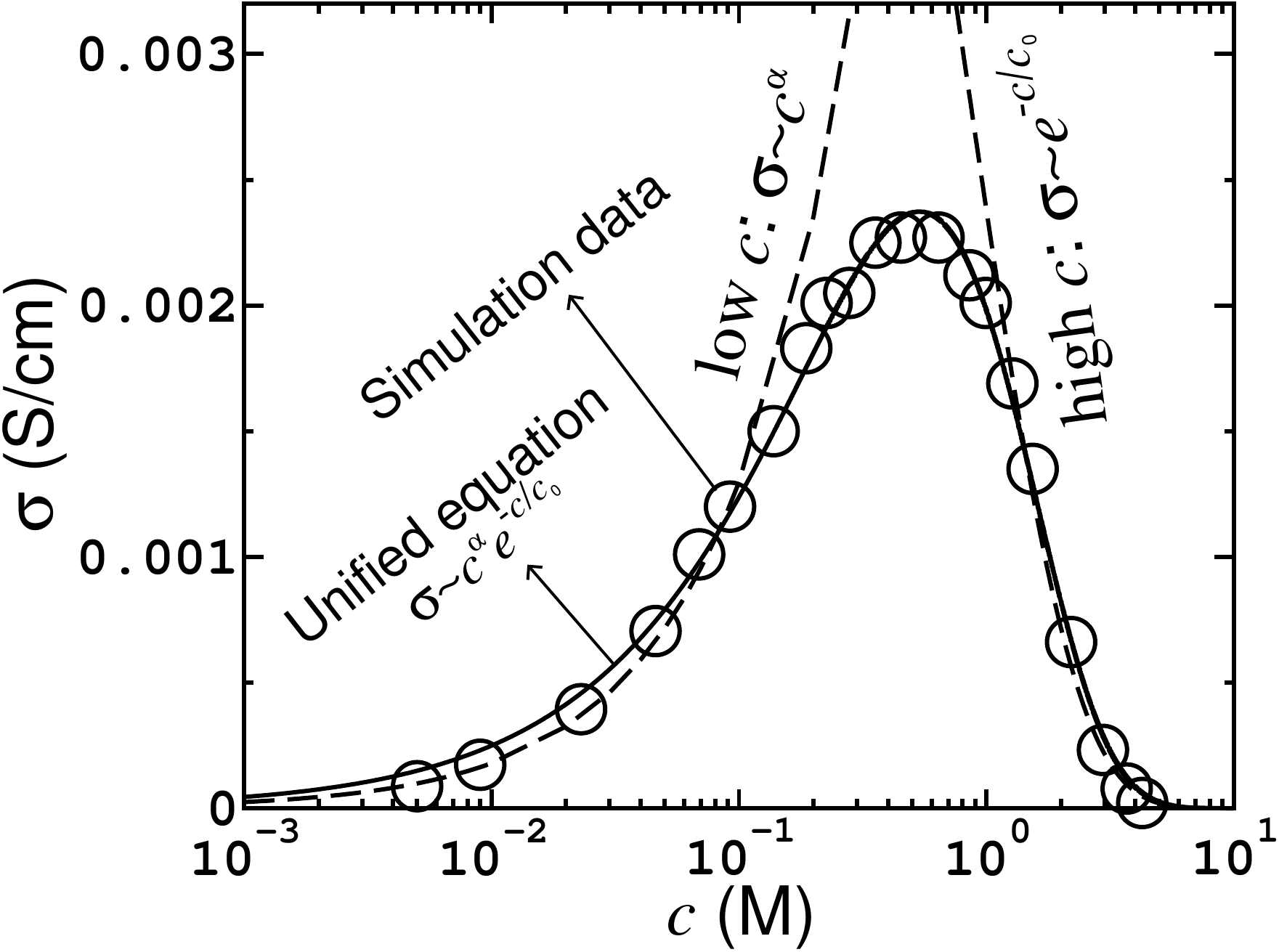}% Here is how to import EPS art
	\caption{Ionic conductivity in EC-LiTFSI electrolytes as a function of salt concentration. The proposed unified equation in this work, $\sigma=k$$c^{\alpha}e^{-c/c_{0}}$, explains $\sigma$ convincingly over a wide range of $c$.}
	\label{fig:figure4}
\end{figure}

In highly concentrated liquid electrolytes composed of LiBF$_4$ and sulfolane, the predominant mechanism of charge transport involves ions moving alongside their respective solvation shells—a process known as vehicular transport.\cite{mukherji2020,bocha2020,brandell,chauvet2023} In contrast, structural diffusion involves ion transport through the dynamic formation and dissociation of ion-pairs, without the ions carrying their full solvation shell.\cite{borodin2006,develop2006} Recent simulations by Balasubramanian and coworkers\cite{mukherji2020} revealed significant dynamic heterogeneity among Li$^+$ ions which was attributed to the caging effects and the occurrence of ion hopping at high salt concentration. Particularly noteworthy is the observation that at higher concentrations, the marked predominance of Li$^+$ ion hopping becomes a clear indicator of its central role in facilitating the efficient transport of Li-ions in the electrolyte. However, Borodin and Smith,\cite{borodin2006,develop2006} demonstrated through MD simulations that in a 1:10 LiTFSI:EC electrolyte, vehicular transport accounts for only approximately 50\% of Li$^+$ transport.

\begin{figure}
	\includegraphics[height=6.5cm,keepaspectratio]{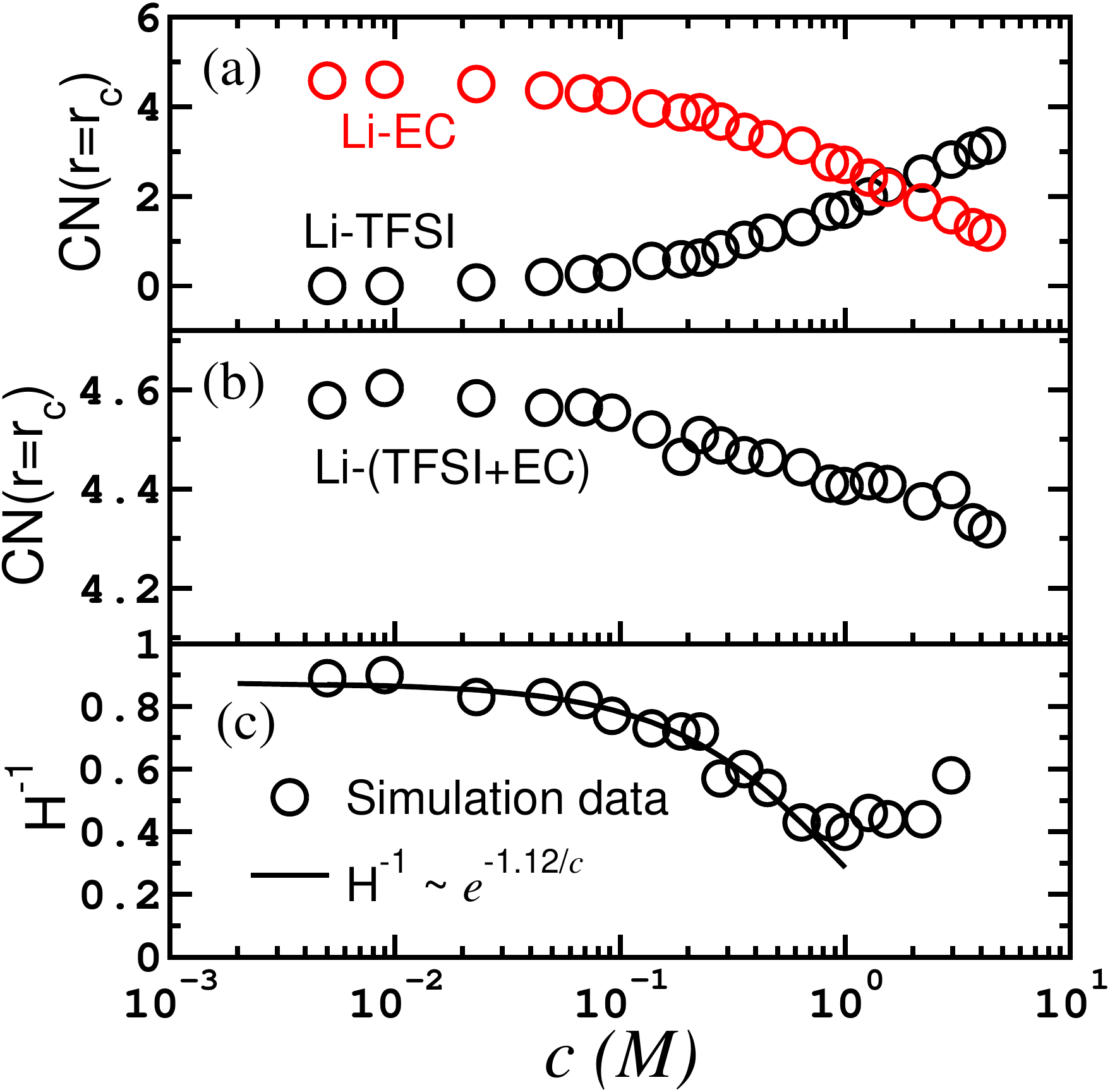}% Here is how to import EPS art
	\caption{(a) Coordinating nitrogen atoms of TFSI$^-$ and oxygen atoms of EC molecules around Li$^+$ ions at the respective cutoff distances, taken as the location where the radial distribution function shows first minimum, (b) Sum of the coordinating TFSI$^-$ and EC molecules around Li$^+$ ions, and (c) The inverse haven ratio or the degree of uncorrelated motion of ions (H$^{-1})$ in the EC-LiTFSI electrolytes as function of $c$. The simulation data was fitted to an exponentially decay function up to 0.64 M salt concentrations, resulting in $H^{-1}=\sigma/\sigma_{NE}=0.875e^{-1.12c}$.}
	\label{fig:figure5}
\end{figure}

We also observed a similar behavior for the concentration corresponding to 1:10 LiTFSI:EC, as supported by the data shown in Fig. \ref{fig:figure5}. The number of EC molecules coordinating around a Li$^+$ ion was found to be under 4, consistent with simulations of Borodin and Smith,\cite{borodin2006} but slightly less than experimental reports\cite{raman1989}. We find that at low $c$, the lithium-ion solvation shell is dominated by EC molecules (see Fig. \ref{fig:figure5}a), indicating vehicular transport. As $c$ increases, the solvation shell shifts toward anion dominance, signaling a transition to structural diffusion through ion-pair formation and dissociation. At salt concentration between 1 and 2 M, we find considerable dominance of both vehicular and structural diffusion mechanisms as assessed based on the criteria involving solvation shell of Li$^+$ ions.

Based on the above discussion, we propose that the ion transport for low $c$ occurs through the vehicular mechanism and for high $c$, the ion transport occurs through the structural relaxation of ion-pairs (i.e., the structural diffusion mechanism). The structural diffusion mechanism at high $c$ is evident from Fig. \ref{fig:figure3} where we observe $\sigma_{NE}\sim\eta^{-1}$ and $\sigma\sim\eta^{-1}$. The proposal is also supported by increased coordination numbers of TFSI$^-$ around Li$^+$ and decrease in coordination numbers of EC around Li$^+$ as the salt concentration increases (see Fig. \ref{fig:figure5}). In electrolytes with considerably higher concentrations, the scenario evolves due to the limited number of solvent molecules per Li$^+$ ion, preventing the formation of a conventional solvation shell. This leads to incomplete solvation of Li$^+$ ions, (see Fig. \ref{fig:figure5}b) triggering low degree of uncorrelated motion of ions (see Fig. \ref{fig:figure5}c) and the formation of larger aggregates (see Section S11). Consequently, in these scenarios, mechanisms such as structural reorganization, ion hopping, and cooperative ion transport modes are likely to make substantial contributions to the overall charge transport process. The structural reorganization mechanism is supported by the fact that $D$ and $\sigma$ correlates well with $\eta$ at high $c$. However, a complete understanding of the ionic conductivity mechanisms across the entire range of $c$ requires further analysis on the structural relaxation timescales of the lithium-ion solvation shell and the lithium-ion hopping dynamics within the electrolyte.\cite{ansarijpcb2024}

In summary, we show that the ionic conductivity increases monotonically at low salt concentrations and displays an extremum before declining with further increase in $c$ despite ionic diffusivities decreasing monotonically with $c$. The increase in ionic conductivity at low $c$ is due to the availability of a high fraction of uncorrelated charge carriers in the electrolyte while the salt induced structural relaxations manifested through the viscosity contributes to the ionic conductivity at high $c$ as $\sigma_{NE}\sim\eta^{-1}$ and $\sigma\sim\eta^{-1}$. We observe two different regimes of salt concentrations dictated by $\eta\sim\tau_c^{1/2}$ and $\eta\sim\tau_c^{1}$, indicating the prominence of two different ion conductivity mechanisms, identified as vehicular and structural diffusion mechanisms. We proposed a unified equation to explain the ionic conductivity dependency on $c$ as $\sigma=kc^{\alpha}e^{-c/c_{0}}$, that fits excellently with our simulation data over the entire range of $c$. The unified equation holds promise in understanding the molecular level origin of ionic conductivity and provides insights into the ion transport mechanisms in a variety of liquid and polymer electrolytes. We propose that the ion transport occurs through the vehicular mechanism below the reach of conductivity maximum, where $\eta\sim\tau_c^{1/2}$, and through the structural relaxation of ion-pairs (i.e, the structural diffusion mechanism) at high salt concentrations, where $\eta\sim\tau_c^{1}$. Overall, the microscopic level ionic transport mechanisms emerged from this work would further assist in the better interpretation and understanding of the experimental results.

\begin{acknowledgments}
	We thank Michael L. Klein for insightful discussions and critical suggestions. We thank IIT Jodhpur for the support provided through the DGX2 and HPC facilities. SM acknowledges support from the SERB International Research Experience Fellowship SIR/2022/000786 provided by the Science and Engineering Research Board, Department of Science and Technology, India. PKJ acknowledges IIT Jodhpur for a Seed Grant (Grant No. I/SEED/PKJ/20220016) and SERB, India for a Core Research Grant (Grant No. CRG/2022/006365). HT and SSPC acknowledge the Ministry of Education (MoE), Govt. of India for the financial support received through fellowship.
\end{acknowledgments}

%\nocite{*}

\bibliography{concentrated_ec_litfsi}% Produces the bibliography via BibTeX.

\end{document}